\documentclass{JHEP3}

\usepackage{amssymb}
\usepackage{amsmath}

\usepackage{graphicx}
\usepackage{graphics}
\usepackage{epsfig}

\usepackage{multirow}

\newcommand{\APJ}{Astrophys.\ J.\/}
 
\newcommand{\CP}{Contemp.\ Phys.\/}
\newcommand{\MNRAS}{Mon.\ Not.\ R.\ Aston.\ Soc.\/}

\newcommand{\PL}{Phys.\ Lett.\/} 
\newcommand{\PR}{Phys.\ Rev.\/} 
\newcommand{\PRL}{Phys.\ Rev.\ Lett.\/} 

\newcommand{\Mnu}{$M_\nu$}

\title{Do WMAP data favor neutrino mass and a coupling between Cold
Dark Matter and Dark Energy$\, $?}

\author{G. La Vacca$^{1,2}$, J.R. Kristiansen$^3$, L.P.L. Colombo$^4$,
R. Mainini$^3$, S. A. Bonometto$^{1,2}$\\
$^1$Physics Department G.~Occhialini, Milano--Bicocca
University, Piazza della Scienza 3, 20126 Milano, Italy\\
$^2$I.N.F.N., Sezione di Milano--Bicocca, Piazza della
Scienza 3, 20126 Milano, Italy\\
$^3$Institute of Theoretical Astrophysics, University of Oslo,
Box 1029, 0315 Oslo, Norway\\
$^4$Department of Physics \& Astronomy, University of
Southern California, Los Angeles, CA 90089-0484\\
E-mail: \email{lavacca@mib.infn.it,j.r.kristiansen@astro.uio.no}}

\abstract{Within the frame of cosmologies where Dark Energy (DE) is a
self--interacting scalar field, we allow for a CDM--DE coupling and
non--zero neutrino masses, simultaneously. In their 0--0 version, {\it
i.e.} in the absence of coupling and neutrino mass, these cosmologies
provide an excellent fit to WMAP, SNIa and deep galaxy sample spectra,
at least as good as $\Lambda$CDM. When the new degrees of freedom are
open, we find that CDM--DE coupling and significant neutrino masses
($\sim 0.1\, $eV per $\nu$ species) are at least as likely as the 0--0
option and, in some cases, even statistically favoured. Results are
obtained by using a Monte Carlo Markov Chain approach.}

\preprint{...}

\begin{document}

\section{Introduction}
In the late 90's Hubble diagrams of SNIa \cite{bib1} became
sufficiently precise to allow the unexpected conclusion that the
cosmic expansion is accelerated. This agreed with fresh Cosmic
Microwave Background (CMB) \cite{bib2} and large scale structure (LSS)
\cite{bib3} data, strongly indicating that the background metric is
{\it spatially flat}, while the matter density parameter $\Omega_{o,m}
\simeq 0.27$ is much below unity. It was then natural to infer that
the rest of the energy budget (up to $\Omega_o \simeq 1$) is
responsible for the cosmic acceleration; dubbed Dark Energy (DE), it
ought to be a smooth non--clustering component, with a state parameter
close to -1.

All data available up to now can be accomodated in a cosmology where
DE has a state parameter $w \equiv -1$, a model equivalent to
introducing Einstein's cosmological constant. This minimal model is
dubbed $\Lambda$CDM (or cosmic concordance) cosmology. However, DE
nature has yet to be properly understood.

A component with $w \equiv -1$ could be {\it false vacuum}. If vacuum
energy does not vanish, its expected density is $m_p^4$ ($m_p$: the
Planck scale), and the measured density implies a {\it fine tuning}
$\sim 1:10^{124}$. But, even referring to the last phase transition,
supposed to occur when the cosmic temperature was $T_{EW} \sim
100~$GeV, the fine tuning is still $\sim 1:10^{56}$.

The {\it coincidence} paradox is even more severe. If one does not
want to indulge to anthropic perspectives, a vacuum energy level just
allowing structure formation, and stopping the process just when it
has completed, can be hardly accepted without a justification.

Alternatives to false vacuum were then proposed, aimed to avoid fine
tuning and coincidence. They mostly bring new parameters to be fitted
to the same data $\Lambda$CDM already fits so nicely. The hope of a
substantial increment of model likelihood was however frustrated,
so that no real new insight into DE physics has yet been gained in
this~way.

A somehow alternative pattern is an unbiased fit of the scale
dependence of DE state parameter $w(a)$ to data.  But, although
restricting to a 2--parameter expression
\begin{equation}
w(a) = w_o + (1-a)w'~,
\label{wa}
\end{equation}
current data hardly do more than fixing a likelihood ellipse
\cite{komatsu}, provocatorily centered on $w_o = -1$, $w'=0$, and up
to now, also this approach has failed to give any new insight into the
DE nature.

\begin{figure}[t!]
\includegraphics[height=7.cm,angle=0]{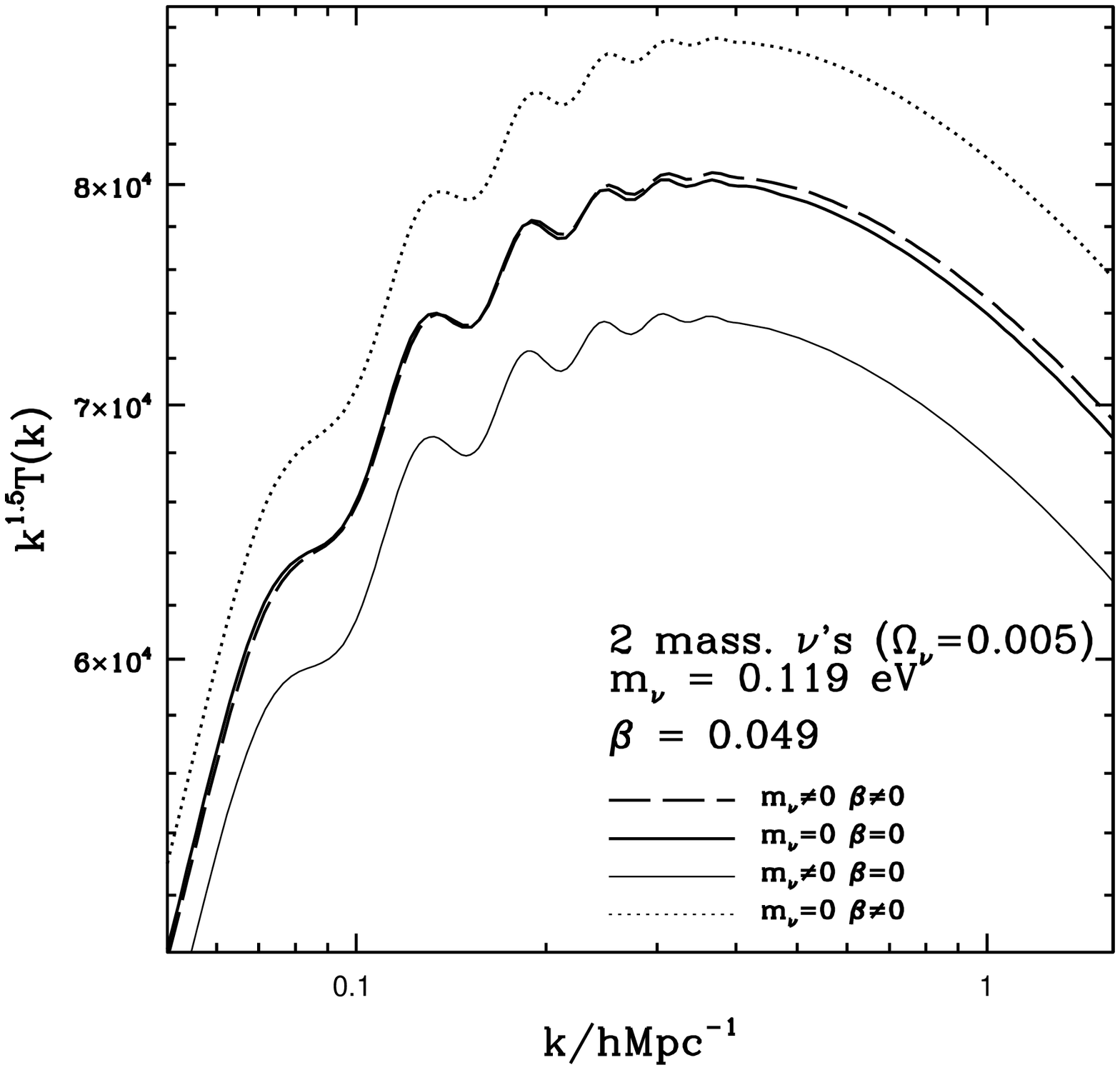}
\hskip .5truecm
\includegraphics[height=7.7cm,angle=0]{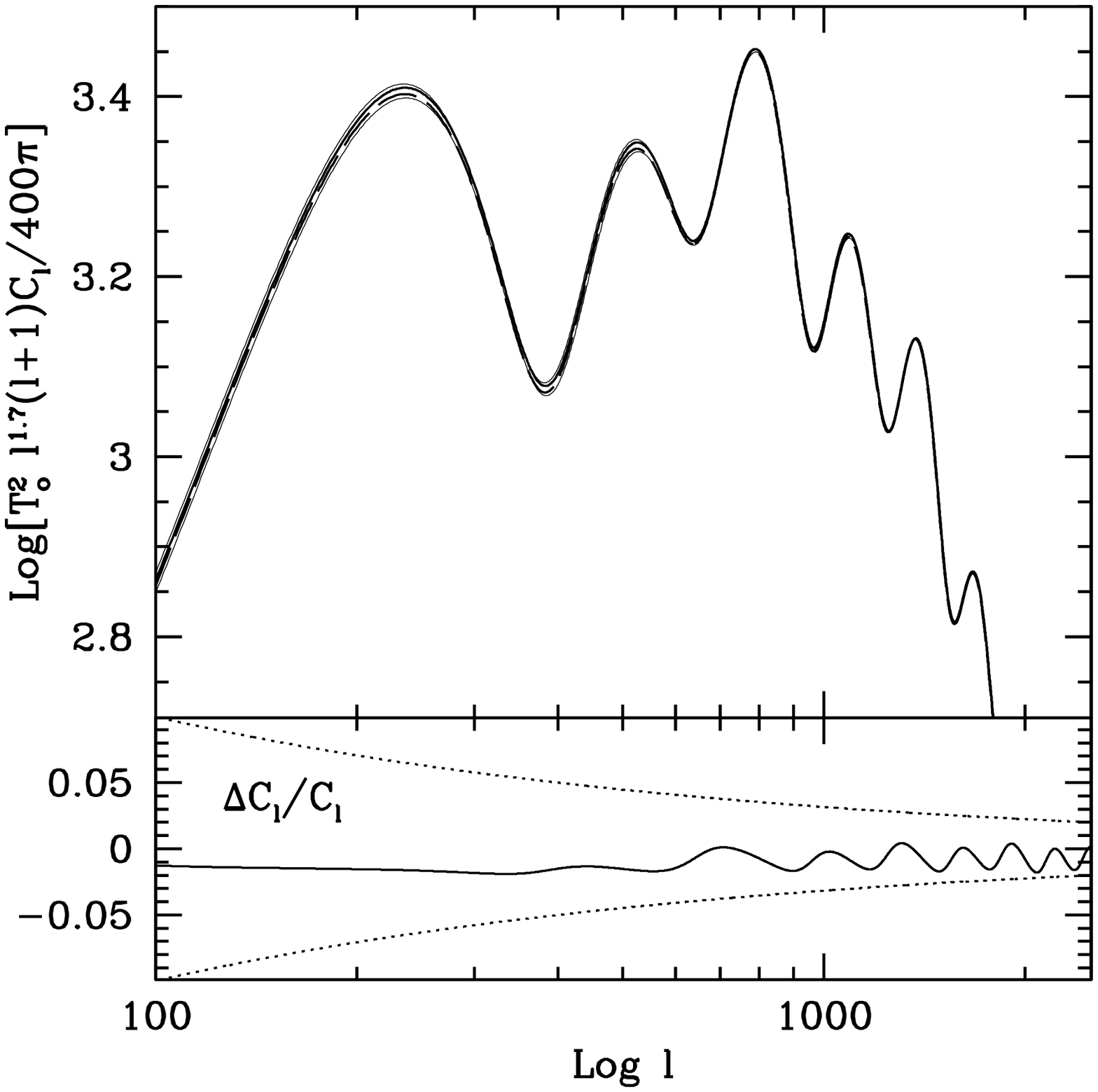}
\caption{Transfer functions (left) and angular anisotropy spectrum
(right) in cosmologies with/without coupling and with/without 2
massive $\nu$'s (00/CM models). Coupling and mass are selected so to
yield an approximate balance. The transfer functions are multiplied by
$k^{1.5}$, to help the reader to distinguish different cases. In the
lower frame of the $C_l$ plot we also give the spectral differences
between 00-- and CM--models, hardly visible in the upper frame. Here
dotted lines represent the cosmic variance interval.}
\label{compense}
\vskip +.1truecm
\end{figure}

In this paper we however keep within the latter approach, but
modifying the parameter budget. We aim to test whether existing data
are already more constraining, once the range of models explored is
different. The option we shall explore is suggested by previous
results based on a Fisher Matrix (FM) analysis \cite{lavacca}. Here
the authors started with observing that spectral distortions due to a
coupling between cold dark matter (CDM) and DE (parametrized by
$\beta$, see below) or to neutrino ($\nu$) masses are essentially
opposite. This holds for both $C_l$ and $P(k)$, the CMB anisotropy
spectrum and the matter fluctuation spectrum (see Figure
\ref{compense} for an example).
Their FM analysis then assumes that a model with vanishing $\nu$
masses and no coupling (the 00--model) has top likelihood, and tests
how far one can go, increasing $\nu$ masses and coupling, keeping
within 1--$\sigma$ or 2--$\sigma$ from the 00--model. When a CDM--DE
coupling is simultaneously allowed, they find that this allows $\nu$
masses well above the limits from the WMAP team \cite{komatsu}, as
quoted in Table \ref{tab:nulim}.
\begin{table}[h]
\centering 
\begin{center}
\begin{tabular}{lcll}
\hline
 && $w = -1$   & $w \neq -1$ \\
\hline\hline
WMAP5        && $< 1.3$~eV  & $< 1.5$~eV   \\
WMAP5+BAO+SN && $< 0.61$~eV & $< 0.66$~eV  \\
\hline
\end{tabular}
\caption{Summary of the 2--$\sigma$ (95\% C.L.) constraints on the sum
of $\nu$ masses \cite{komatsu}, from WMAP 5-year, Baryonic Acoustic
Oscillations and SuperNova data sets.}
\label{tab:nulim} 
\end{center}
\end{table}

In this work we shall therefore perform a full Monte Carlo Markov
Chain (MCMC) analysis of available data, simultaneously opening the
degrees of freedom of $\nu$ mass and CDM--DE coupling. 

Let us remind that CDM--DE coupling \cite{coupling} eases the
coincidence problem. Within $\Lambda$CDM, at $z \sim 10^3$ the
DE--matter density ratio is $\sim 10^{-9}$. With coupling, instead, it
could already exceed $10^{-2}$. With (quasi--)vanishing $\nu$ masses,
however, such coupled models clash against data, setting a limit
$\beta \lesssim 0.075$ \cite{coupling,claudia,maccio}.

In turn, at least one $\nu$ mass eigenstate or, possibly, two of them
exceed $\simeq 0.055~$eV (direct or inverse hierarchy).  This follows
from solar \cite{solar} and reactor \cite{reactor} neutrino
experiments, yielding $\Delta m_{1,2}^2 \simeq 8 \times
10^{-5}$eV$^2$, and atmospheric \cite{atmo} and accelerator beam
\cite{beam} experiments yielding $\Delta m_{2,3}^2 \simeq 3 \times
10^{-3}$eV$^2$. However, the neutrino oscillation experiments do not
provide us with any information on the absolute scale of $\nu$ masses.

Cosmology is sensitive to $\nu$ masses. Already in 1984 Valdarnini \&
Bonometto \cite{bono1} derived the transfer function for mixed DM
models, with DM partially made of massive $\nu$'s. Mixed models were
widely tested in the Nineties. $\nu$'s then filled the apparently
unescapable gap between $\Omega_{om}$ and $\Omega_o~.$ That gap is now
neatly defined and filled by DE, as already outlined. But $\nu$'s
becoming non-relativistic cause so strong spectral distorsions that
even a small contribution to the density budget from them can be
tested (for a thorough review on effects of massive $\nu$'s on
cosmological observables, see \cite{lesgourgues:2006}). This gives
the cosmological limits on the absolute scale of $\nu$ masses in Table
1, an order of magnitude more stringent than limits from tritium
$\beta$--decay experiments.

The ambitious aim of this paper is then to show that, as both limits
can be substantially relaxed, (mildly--)mixed coupled models could
really be an alternative to the minimal $\Lambda$CDM cosmology with
the advantage of easing both fine tuning and coincidence paradox.

The plan of the paper is as follows: In Section \ref{sec:potentials}
we outline the expressions and potentials related to the interacting
CDM--DE fluid. In Section \ref{sec:coupling} we review coupled DE
(cDE). The data sets and statistically methods used are presented in
Section \ref{sec:dandm}. In Section \ref{sec:results} we present our
results. Finally we summarize our findings and conclude in Section
\ref{sec:summary}.

\section{Self--interaction potentials} \label{sec:potentials}

\begin{figure}[t]
\begin{center}
\includegraphics[height=7.cm,angle=0]{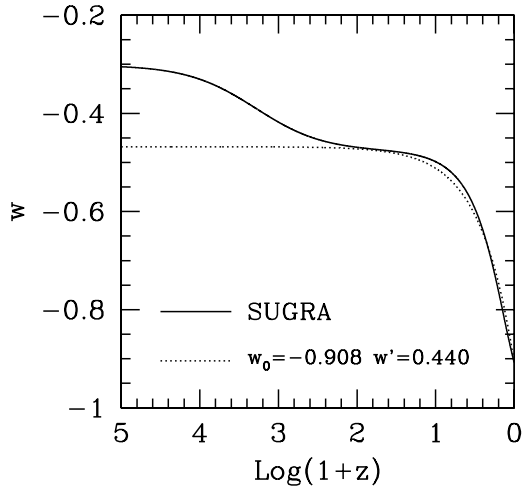}
\end{center}
\caption[]{State parameter in models with $w(a)$ given by
eq.~(\ref{wa}) and in a uncoupled SUGRA model with $\Lambda = 0.1\,
$GeV. $w_o$ and $w'$ values (in the frame) selected to yield a
behavior close to SUGRA, by requiring similar high--$z$ plateau and
$w(0)$. Although renouncing to a full coincidence at $z=0$, the fast
variablility of $w(a) $ in uncoupled SUGRA cannot be met by any
polinomial $w(a)$.  The plot is for $h=0.7$, $\Omega_b=0.046$,
$\Omega_c= 0.209$.}
\label{SU}
\end{figure}

We shall assume that DE is a scalar field $\phi$, self--interacting
through the potentials
$$
V(\phi) = \Lambda^{\alpha+4}/\phi^\alpha  ~~~~~~~~~~~~~~~~~~
\eqno{RP~~~~~~~~}
$$
or
$$
V(\phi) = (\Lambda^{\alpha+4}/\phi^\alpha) \exp(4\pi\, \phi^2/m_p^2)
\eqno{SUGRA~~~~~}
$$
admitting tracker solutions. Uncoupled RP \cite{RP88} yields a slowly
varying $w(a)$ state parameter, steadily below -0.85 for
\begin{figure}[t]
\begin{center}
\includegraphics[height=7.cm,angle=0]{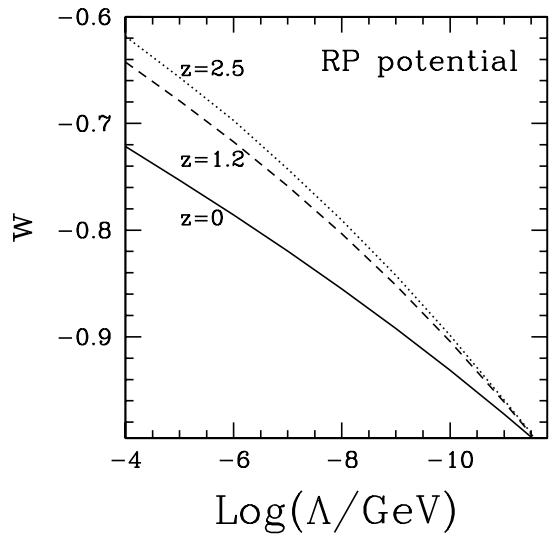}
\end{center}
\caption{State parameter and its variation in uncoupled RP models.
The plot is for $h=0.7$, $\Omega_b=0.046$, $\Omega_c= 0.209$.}
\label{RP}
\end{figure}
$\Lambda/{\rm GeV} \lesssim 10^{-9}$ (see Figure \ref{RP}). On the
contrary, uncoupled SUGRA \cite{SUGRA} yields a fastly varying $w(a)$,
even faster than any expression (\ref{wa}), as is shown in Figure
\ref{SU}.  Coupling is however an essential feature that we shall be
considering and, in the next Section, we shall see how it modifies
these behaviors.

Independently of the presence of coupling, for any choice of $\Lambda$
and $\alpha$ these potentials yield a precise DE density parameter
$\Omega_{de}$. Here we use $\Lambda$ and $\Omega_{de}$ as free
parameters; the related $\alpha$ value is then suitably fixed.

Both RP and SUGRA potentials were initially introduced to ease the
fine tuning problem. In fact, the present DE density can be tuned on
today's CMB temperature $T_o$, reading $\rho_{o,de} \sim (10\,
T_o)^4$; tracker solutions then require a present DE field $\phi_o
\sim m_p$ so that an approximation $V(\phi_o) \sim \rho_{o,de}$ yields
$m_p^\alpha (10\, T_o)^4 \sim \Lambda^{4+\alpha}$. For reasonable
$\alpha$'s, $\Lambda$ is therefore of the order of the energy scale of
SUSY breaking or EW transition.

Fits to data however show that uncoupled RP fails to meet CMB
anisotropy data, unless the energy scale $\Lambda $ stands below $\sim
10^{-9}{\rm GeV}$ \cite{bib6} (previous approximate relations indicate
then $\alpha \sim 0.5$); this is coherent with the fact that $w(a)$ is
then steadily below -0.85 ($ \sim 95\, \% $~C.L.) as established in
\cite{komatsu} for constant $w(a)$ .

In the case of uncoupled SUGRA, however, while the best fit still
leads to low $\Lambda$'s, a value $\Lambda \sim \cal O$$(0.1\, {\rm
GeV})$ is within $\sim 2\, \sigma$'s from the best--fit model. Let
alone the SUSY and EW scale, this is still an energy close to the
confinement scale. Coupling however causes quantitative modifications
to these behaviors, that we shall see in the next Sections.

Quite in general, however, the very fact that the natural
representation to describe DE is not the one where particle numbers
are diagonal, is related to the smallness of the mass of the
quanta. Such low particle masses, although being a natural consequence
of the potential expressions, are considered a hidden fine--tuning by
some researchers.

\section{CDM--DE coupling}
\label{sec:coupling}

In the cDE scenario, as for dynamical DE, the scalar
field $\phi$ yields a cosmic component unsuitable to clustering and
showing a negative pressure in the present epoch.  Its energy density
and pressure however read
\begin{equation}
\rho = \rho_k + V(\phi)~,~~~ p = \rho_k - V(\phi)~,~~~~~~~~~~~
{\rm with}~~\rho_k = \dot \phi^2/2a^2~;
\label{rhop}
\end{equation}
dots indicate differentiation in respect to $\tau$ (conformal time),
the background metrics being
\begin{equation}
ds^2 = a^2(\tau) \left[ d\tau^2 - d\lambda^2 \right]
~~~{\rm with}~~~d\lambda^2 = dr^2 + r^2(d\theta^2 + cos^2 \theta
\, d\phi^2)~.
\label{metric}
\end{equation}
These expressions show that two regimes are possible. If $\rho_k \gg
V$, the DE state parameter approaches +1 ({\it stiff matter}) so that
DE energy density rapidy dilutes during expansion ($\rho \propto
a^{-6}$). In the opposite case $V \gg \rho_k$, the state parameter
approaches --1 and DE is suitable to explain the observed cosmic
acceleration.

Let us now consider the possibility that DE is coupled to other
components. Interactions with baryons, constrained by observational
limits on violations of the equivalence principle (see, {\it e.g.}
\cite{darmour1}), are almost fully excluded. Similar constraints,
however, do not exist for DE--CDM interactions. The only constraints
then derive from cosmological data.

The simplest possible coupling is a linear one.  It can be formally
obtained by performing a conformal transformation of Brans--Dicke
theory (see, {\it e.g.}, \cite{brans}), where gravity is modified by
adding a $\phi R$ term to the GR Lagrangian ($R$: Ricci scalar).
Coupling causes an energy transfer between CDM and DE, so allowing DE
to have a non--negligible density even when its state parameter is
$\sim +1$. However, $\rho_k$ being then dominant, the transfered
energy is soon diluted.

A so--called $\phi$--matter dominated period then occurs when, because
of the power leaking towards DE, CDM density declines more rapidly
than $a^{-3}$.  The increase of $\phi$ then brings it to approach
$m_p$ and $V(\phi) $ to exceed $\rho_k$. DE dilution then stops and DE
eventually exceeds DM density.

The overall picture is however quite natural. All tenable cosmological
models do require a dark sector, split into two components with
different state equations. The fact that their interactions with
baryonic matter is just gravitational, leads to requiring that
\begin{equation}
\label{conti0}
T^{(c)~\mu}_{~~~~\nu;\mu} + T^{(de)~\mu}_{~~~\, ~~\nu;\mu} = 0
\end{equation}
(here $T^{(c,de)}_{\mu\nu}$ are the stress--energy tensors of CDM and
DE, let their traces read $T^{(c,de)}$), while the assumption that CDM
and DE are two separate fluids leads to take $C \equiv 0$ in the
relations
\begin{align}
T^{(de)~\mu}_{~~~\, ~~\nu;\mu} =& +C T^{(c)} \phi_{,\nu}\\
T^{(c)~\mu}_{~~~~\nu;\mu} =&- C T^{(c)} \phi_{,\nu}~,
\end{align}
describing the most general form of linear coupling (incidentally,
this shows why DE cannot be linearly coupled to any component
with vanishing stress--energy tensor trace).
\begin{figure}[!t]
\vskip .5truecm
\begin{center}
\includegraphics[height=9.cm,angle=0]{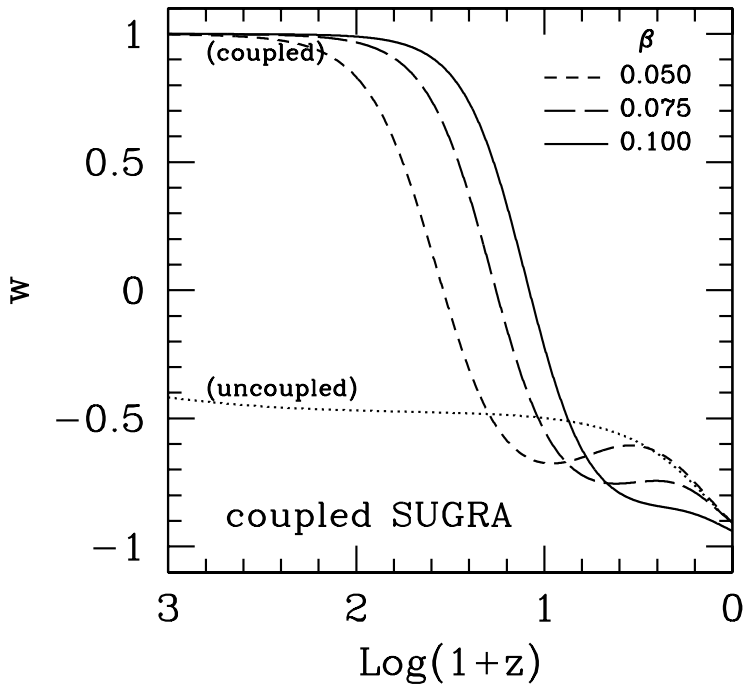}
\end{center}
\caption[]{State parameter in coupled SUGRA model with $\Lambda
= 0.1\, $GeV, $h=0.7$, $\Omega_b=0.046$, $\Omega_c= 0.209$, as
in Figure \ref{SU}. The uncoupled behavior given there is
reported also here.}
\label{SUwc}
\end{figure}

Assuming two separate fluids is clearly an extra assumption and, if we
allow for $C \neq 0$, when the metric is (\ref{metric}), these
equations yield
\begin{align}
\ddot \phi + 2 {\dot a \over a} \dot \phi + a^2 V'_\phi =& +C a^2 \rho_c \\
\dot \rho_c + 3 {\dot a \over a} \rho_c =&  -C \rho_c \dot \phi
\end{align}
$\rho_c$ being CDM energy density. 

General covariance requires $C$ to be a constant or to evolve as a
function of $\phi$ itself.  Here, instead of $C$, we shall mostly use
the adimensional parameter
\begin{equation}
\label{bdef}
\beta = (3/16\pi)^{1/2} m_p C
\end{equation}
and consider constant $\beta$ values $\cal O$$(0.1)$ (corresponding to
$C \sim 1/2m_p$) which, as we shall see, meet observational data.
\begin{figure}[!t]
\begin{center}
\includegraphics[height=12.cm,angle=0]{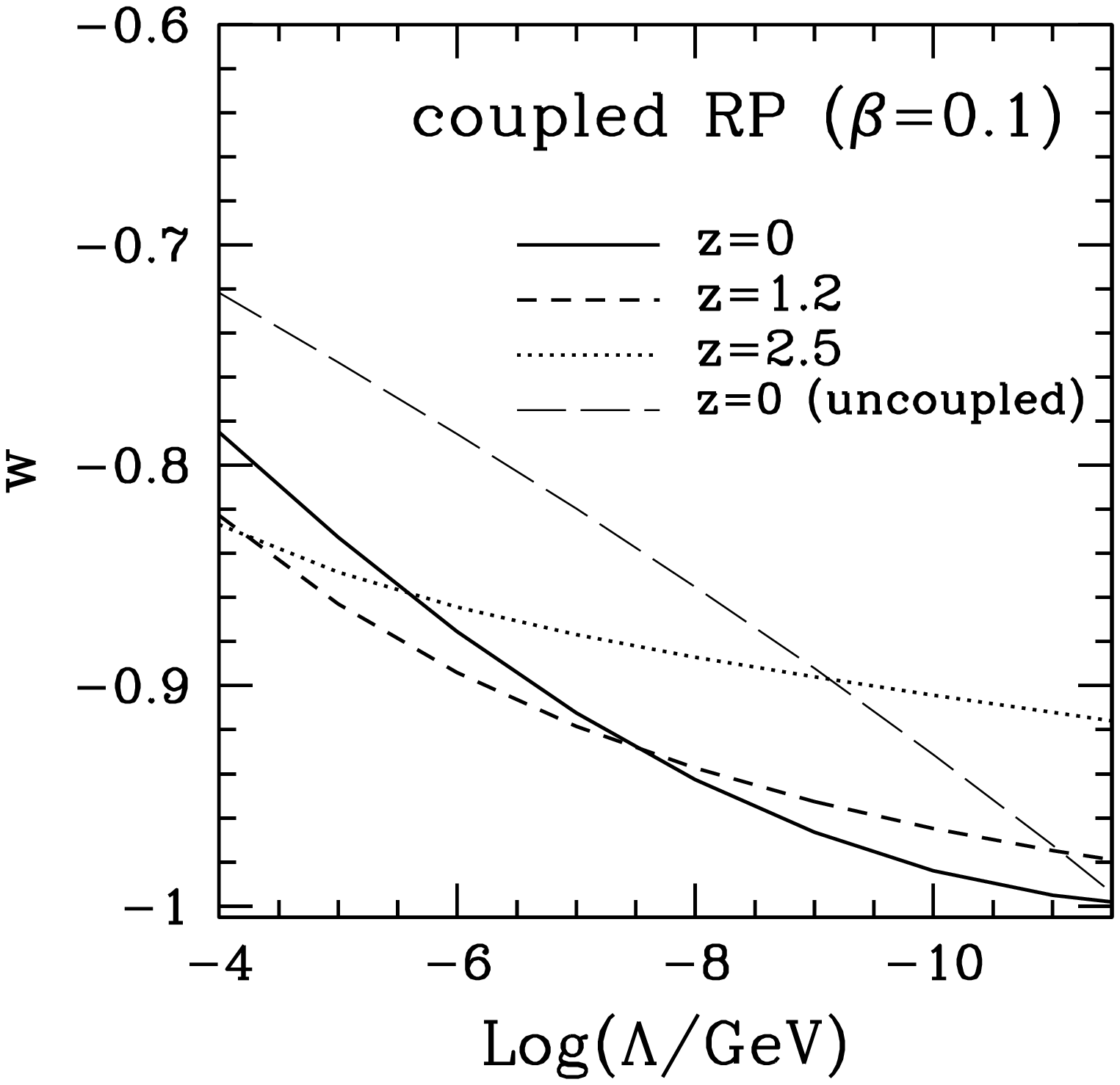}
\end{center}
\vskip-2.4truecm
\caption[]{State parameter in coupled RP for various $\Lambda$ values,
in the case $\beta = 0.1\, $. Here $h=0.7$, $\Omega_b=0.046$,
$\Omega_c= 0.209$, as in Figure \ref{RP}. The uncoupled behavior at
$z=0$ given there is reported also here.}
\label{cRP}
\end{figure}

The above dynamics naturally causes a $w(a)$ behavior significantly
different from the uncoupled case. In Figure \ref{SUwc} we plot the
$z$ dependence of the DE state parameter for the SUGRA potential.  For
the sake of comparison the uncoupled behavior is also shown.

In the Figure, the transition from a {\it stiff--matter} behavior to
$w \sim -1$ is clearly shown. Such transition is later for greater
coupling. At low $z$, for low $\beta$'s, the uncoupled behavior is
reapproached and, for $\beta$ as low as 0.05~, the dependence of $w$
on $z$, when approaching $z=0,$ is even steeper than in the
uncoupled case.  For greater couplings, however, the behavior
gradually softens and, for $\beta \simeq 0.1$, the effective value of
$w$ at low $z$ is systematically closer to $\Lambda$CDM than in the
uncoupled case.

Figure \ref{SUwc} also shows that the scale dependence of the state
parameter exhibits a peculiar feature, even in the presence of mild
couplings ({\it e.g.}, for $\beta = 0.05$), shifting from negative to
positive values in a potentially observable redshift range. This
behavior is not restricted to SUGRA potentials; it holds for RP and
other potentials as well. Even approximate measurements of the scale
dependence of the state parameter, extending up to $z \sim 10$--30,
should therefore be able to set direct limits to the coupling
intensity.

The behaviors are qualitatively similar in the case of a RP
potential. Figure \ref{cRP}, similar to Figure \ref{RP}, includes a
coupling $\beta = 0.1$. It shows, first of all, that $w$ still keeps
fairly constant values. Furthermore, at low $z$, the state parameter
$w$ is not necessarily increasing with $z$, as it may occur also for
the SUGRA potential and is shown in Figure \ref{SUwc} (for $\Lambda =
10^{-1}~$GeV), {\it e.g.} when $\beta = 0.05$. As for SUGRA, at low
$z$, $w(a)$ is mostly smaller in the presence of coupling. The $w =
-0.85$ limit is bypassed here around $\Lambda = 10^{-5}~$GeV. We
therefore expect that a fit of coupled RP with data will be already
fair for $\Lambda$ values greater and physically more significant.

Before concluding this Section, it may be worth defining the coupling
function $f(\phi)$, through the relation
\begin{equation}
C(\phi) = {d \log f \over d\phi}
\end{equation}
so that CDM energy density scales according to
\begin{equation}
\label{cevol}
\rho_c(a) = \rho_{o,c} (a_o/a)^3 f(\phi)~.
\end{equation}
Then, if we set $\bar V = V + \rho_c~,$ the $\phi$ eq.~of motion takes
the (standard) form,
\begin{equation}
\ddot \phi + 2 {\dot a \over a} \dot \phi + a^2 \bar V'_\phi = 0~,
\end{equation}
as though CDM and DE were decoupled, once the {\it effective}
potential $\bar V$ is used.

The CDM evolution (\ref{cevol}) is then faster than in the absence of
coupling. In turn, the effective behavior, obtainable by using the
potential $\bar V$, mimics a {\it phantom}--like state equation,
yielding a DE density increase with $a$, as we would find for $w <
-1~.$

This makes reasonable that neutrino mass limits are relaxed in a cDE
context, as they are in the presence of phantom DE. This option,
however, does not lead to requiring unconventional physics. On the
contrary, by assuming quite a general behavior within the dark sector
and adding the ingredient of $\nu$ mass, if we are naturally led to
high $\beta$ values, this will also ease the coincidence problem.

\section{Data and methods} \label{sec:dandm}

In this Section we present data sets and methods applied in our
analysis. 

To constrain the CMB power spectra, we use the 5 year data from the
WMAP satellite \cite{wmap} (WMAP5). We use both temperature and
polarization data, and calculate the likelihood of our models using
the Fortran 90 code provided by the WMAP
team\footnote{http://lambda.gsfc.nasa.gov; version v3}.

For the matter power spectrum, we use the results from the 2dF galaxy
redshift survey \cite{2df}. Constraints on the recent expansion
history of the Universe is also given by the SN1a observations from
the Supernova Legacy Survey (SNLS) \cite{SNLS}.

In some cases we also apply additional priors on the Hubble
parameter and the baryon density of the Universe. The prior on the
Hubble parameter of $h = 0.72 \pm 0.08$ is taken from the Hubble Space
Telescope (HST) Key Project \cite{HST}. Analysis of the Big Bang
Nucleosynthesis \cite{BBN} gives a prior on the baryon content of
$\Omega_b h^2 = 0.022 \pm 0.002$.

Using many data sets will of course give stronger parameter
constraints, but every additional data set might also introduce new
systematics. Accordingly, we considered three different combinations
of data sets in our analysis; (i) WMAP5 data only, (ii) WMAP + 2dF,
(iii) WMAP + 2dF + SNLS + HST and BBN priors (henceforth named ``all
data'').

For the MCMC analysis we use the publicly available code CosmoMC
\cite{cosmomc} which, in turn, uses theoretical power spectra computed
by a modified version of CAMB \cite{camb}. CMB lensing effects are
included, to ensure accurate results when comparing with CMB
polarization data. In the MCMC analysis we use the following set of
basic parameters: \{ $\omega_b, \omega_{c}, \theta, \tau, n_s, \ln
10^{10} A_s, \Lambda, \beta, M_\nu $ \}. Here: $\omega_{b,c}$ are the
physical baryon and cold dark matter density parameters,
$\omega_{b,c}=\Omega_{b,c} h^2$, where $h$ is the dimensionless Hubble
parameter; $\theta$ is the ratio of the sound horizon to the angular
diameter distance at recombination; $\tau$ is the optical depth to
reionization; $n_s$ is the scalar spectral index; $A_s$ denotes the
amplitude of the scalar fluctuations at a scale of $k=0.002
\textrm{Mpc}^{-1}$. In addition we include the sum of $\nu$ masses,
$M_\nu = \Sigma m_\nu$, assuming 3 equal mass $\nu$'s. Let us outline
that $\Lambda$ denotes the energy scale in DE potentials, while
$\beta$ is the coupling parameter between CDM and DE.

All parameters in the braces are given flat priors, unless otherwise
is stated explicitly. In our MCMC analysis we also marginalize over
the Sunyaev Zel'dovich amplitude, as is also done by the WMAP team in
their analysis. We assume that the Universe is spatially flat. The
chains are run on the Titan cluster at the University of Oslo.

In addition to MCMC runs with the full parameter set presented above,
we have also repeated the analysis without \Mnu~ or without $\Lambda$
and $\beta$, to be able to compare the effects of the different
extensions of the parameter space.

\section{Results} 
\label{sec:results}

In this Section we shall discuss the results of the above MCMC runs.
In the cases already considered in the literature, we reobtain
standard results and likelihood $\cal L$ values. 
In Table \ref{tab:comparison} we report best fit values and
1--$\sigma$ errors for different models. 

When including the $\beta$, $\Lambda$ and $M_\nu$ degrees of freedom
we see that all the other parameters stay within 1$\sigma$ shifts from
their previous mean values. As could be expected, the error bars
increase on some of the parameters, especially $\omega_{c}$.

\begin{figure}[h!]
\vskip-.1truecm
\begin{center}
\includegraphics[height=9.2cm,angle=0]{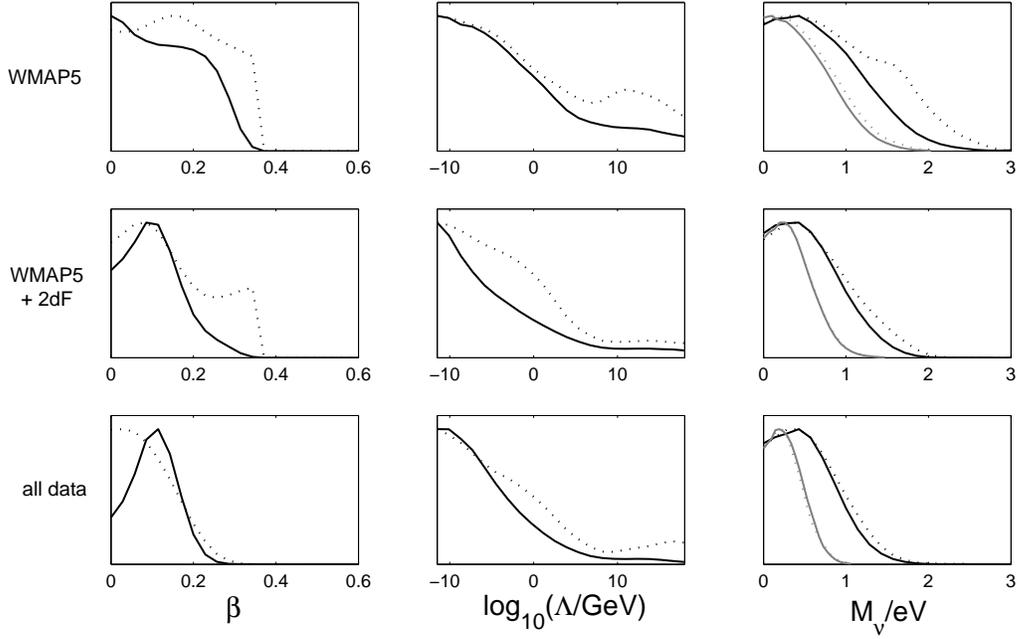}
\end{center}
\vskip-.5truecm
\caption{Marginalized (solid black line) and average (dotted
black line) likelihood of cosmological parameters in SUGRA models. For
\Mnu we also show the corresponding likelihood distributions obtained
in the case of a standard $\Lambda$CDM+\Mnu model (gray lines).}
\label{mnuall}
\end{figure}

\begin{figure}[h!]
\vskip-.1truecm
\begin{center}
\includegraphics[height=9.2cm,angle=0]{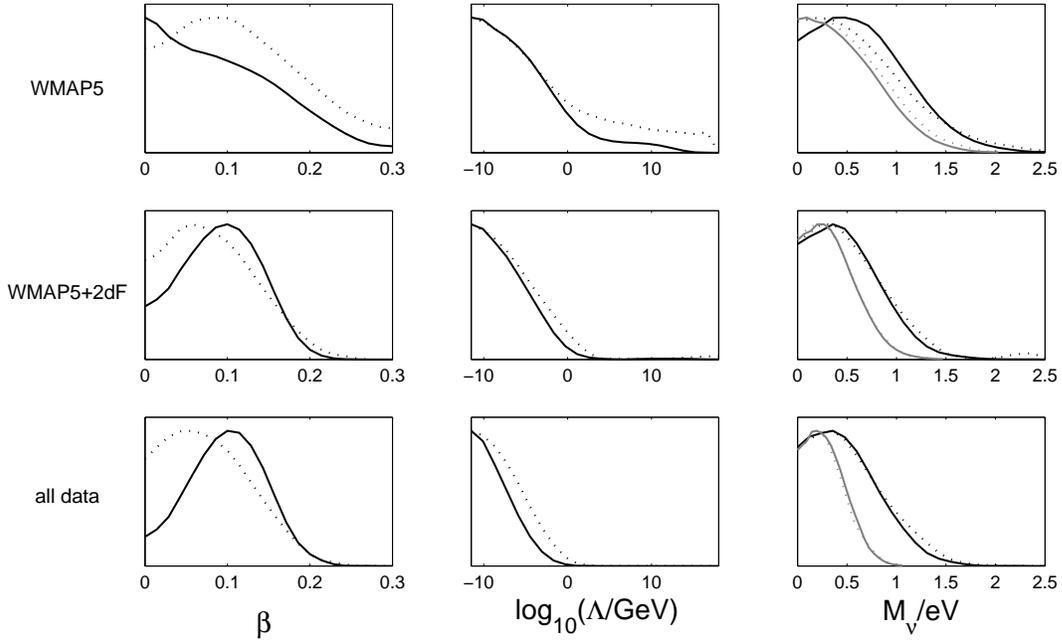}
\end{center}
\vskip-.5truecm
\caption{Marginalized (solid line) and average (dotted line)
likelihood of cosmological parameters in RP models, taking into
account the whole set of available data.}
\label{Rmnuall}
\end{figure}

\begin{table}[!t]
\begin{center}
\begin{tabular}{clllll}
\hline
\multirow{2}{*}{\rm Parameter}&\multicolumn{2}{c}{$\Lambda$CDM + $\nu$'s} & $w =$ const. 
  & cRP + $\nu$'s & cSUGRA + $\nu$'s \\
& WMAP only & all data & all data & all data & all data \\
\hline
\\
\multirow{2}{*}{$10^2\, \omega_{b}$}
  &  2.244       &  2.258       &  2.247        &  2.260       &  2.260      \\
  &  $\pm$ 0.066 &  $\pm$ 0.061 &  $\pm$ 0.062  &  $\pm$ 0.061 &  $\pm$ 0.065\\
\\
\multirow{2}{*}{$\omega_{c}$}
  &  0.1156       &    0.1098    & 0.1132       & 0.1039       & 0.1042       \\
  &  $\pm$ 0.0078 & $\pm$ 0.0040 & $\pm$ 0.0069 & $\pm$ 0.0062 & $\pm$ 0.0084 \\
\\
\multirow{2}{*}{$10^2\theta$}
  &  1.0401       &  1.0401       & 1.0402       & 1.0401       & 1.0406       \\
  &  $\pm$ 0.0030 &  $\pm$ 0.0030 & $\pm$ 0.0030 & $\pm$ 0.0029 & $\pm$ 0.0030 \\
\\
\multirow{2}{*}{$\tau$}
  & 0.085       & 0.087       & 0.085       & 0.087       & 0.088       \\   
  & $\pm$ 0.017 & $\pm$ 0.017 & $\pm$ 0.017 & $\pm$ 0.016 & $\pm$ 0.017 \\
\\
$M_\nu$ (eV)  & \multirow{2}{*}{$<$~1.20} & \multirow{2}{*}{$<$~0.66 } & \multirow{2}{*}{$<$~0.94}
 & \multirow{2}{*}{$<$~1.13} & \multirow{2}{*}{$<$~1.17} \\   
(95\% C.L.)&\\
\\
$\beta$  & \multirow{2}{*}{$-$} & \multirow{2}{*}{$-$} & \multirow{2}{*}{$-$}
 & \multirow{2}{*}{$<$0.17} & \multirow{2}{*}{$<$0.18} \\   
(95\% C.L.)&\\
\\
$\log_{10}(\Lambda/\textrm{GeV})$  & \multirow{2}{*}{$-$} & \multirow{2}{*}{$-$} & \multirow{2}{*}{$-$}
 & \multirow{2}{*}{$<$~-4.2} & \multirow{2}{*}{$<$~6.3} \\   
(95\% C.L.)&\\
\\
\multirow{2}{*}{$n_s$}
  & 0.955       & 0.962       & 0.958       & 0.969       & 0.970       \\
  & $\pm$ 0.017 & $\pm$ 0.014 & $\pm$ 0.015 & $\pm$ 0.015 & $\pm$ 0.018 \\
\\
\multirow{2}{*}{{\rm ln}$(10^{10}A_s)$}
  & 3.053       & 3.045       & 3.049       & 3.055       & 3.057       \\
  & $\pm$ 0.043 & $\pm$ 0.040 & $\pm$ 0.040 & $\pm$ 0.040 & $\pm$ 0.041 \\
\\
\multirow{2}{*}{$\sigma_8$}
 & 0.691       & 0.713       & 0.711       & 0.723      & 0.717       \\
 & $\pm$ 0.075 & $\pm$ 0.056 & $\pm$ 0.059 & $\pm$ 0.062 & $\pm$ 0.069 \\
\\
\multirow{2}{*}{$H_o$ {\rm (km/s/Mpc)}}
  & 67.0       & 70.1      & 69.7      & 71.8      & 71.9      \\
  & $\pm$  4.4 & $\pm$ 2.1 & $\pm$ 2.2 & $\pm$ 2.5 & $\pm$ 2.7 \\
\\
\hline
-2 ln($\cal L$) & 1329.39 & 1407.25   &        1407.38  &       1407.44      &      1407.33\\
\end{tabular}
\caption{Best fit values and 1--$\sigma$ error bars.  In all fits we
allow for $\nu$ masses. The first 9 lines concern primary
parameters. Only upper limits on \Mnu, $\beta$ and $\Lambda$~are
shown. These variables are discussed more thoroughly in forthcoming
2--D plots. Likelihood values are almost model independent.}
\label{tab:comparison}
\end{center}
\end{table}
The most intriguing part of our outputs however concerns \Mnu ~and
$\beta$. Already from Table \ref{tab:comparison} one appreciates that
the upper limit on $\nu$ masses has doubled. This can also be seen in
the Figures \ref{mnuall} and \ref{Rmnuall}, where we show marginalized
and average likelihood distributions on the most specific parameters
of this work: $\beta$, $\log(\Lambda/{\rm GeV})$ and \Mnu.

Figures from the top to the bottom line refer to fits based on
increasingly wide data samples. Taking WMAP5 data only, the
marginalized likelihood distribution gives just a 95$\, \%$ C.L. upper
limit $\beta < 0.28$ and $\beta<0.23$ for SUGRA and RP potentials,
respectively. When low--$z$ data are simultaneously considered, the
marginalized likelihood distributions exhibit significant maxima,
which could be naively interpreted as a $\sim 2\, \sigma$ detection of
$\beta$. A physical reason of the effect can be found in the actual
tension between $\omega_{c}$ mean values, obtained from pure WMAP data
or including low--$z$ data, visible also in Table
\ref{tab:comparison}. A significant DE amount at high $z$ could reduce
there the required value of $\omega_{c}$ by 1-2$\, \%$, and set it
within 1$\, \sigma$ from its all--data value. 

In turn, models with $\beta$ are somehow more likely, although one
must be cautious on this point, where non--Gaussian behaviors become
important, in the presence however of tiny signals. It is true, in
fact, that the peak of the marginalized likelihood, in some cases,
exceeds 0--0 models by almost 2 $\sigma$'s. But the peak is not so
pronounced among average likelihood values: here the top value is
atmost double, in respect to 0--0 models.  Finally, if we consider the
top overall likelihood, it does not exceed $\Lambda$CDM
likelihood. This seems to conflict with the fact that the likelihood
of 0--0 models in RP and SUGRA cosmologies equals the likelihood of
$w=-1$, among $w={\rm const}$ cosmologies. But, of course, the
likelihood of individual models can easily behave differently from
averaged likelihoods.

As far as the $\Lambda$ scale is concerned, the 2--$\sigma$ upper
limit on $\log_{10} \Lambda$ is much softened compared to studies of
uncoupled models. The usual upper limit $\sim 0.1\, $GeV, for SUGRA
models, shifts now above $10^6$GeV. Something similar occurs for the
scale $\Lambda$ in RP models, which is now consistent with data,
within $\sim 2\, \sigma$'s, up to $\sim 10^{-4}$GeV. This had been
somehow predicted from Figure \ref{cRP}, as outlined there.

Finally, in the \Mnu ~distributions, the gray line shows the
distributions in the absence of coupling, and allows to appreciate why
the 95$\, \%$ upper limits in Table \ref{tab:comparison} have almost
doubled. The maxima in the likelihood distributions are far less
accentuated here, than for $\beta$. They are somehow stronger in the
SUGRA than in the RP case. It is however clear that there is no
hint of $\nu$--mass detection in these plots.

In all these plots there are discrepancies between average and
marginalized likelihood distributions. They are particularly 
relevant as far as $\beta$ is concerned. 

Such discrepancies, first of all, are a safe indication of
non--Gaussian distributions. In the case of $\beta$ they can be better
undestood in association with some Figure herebelow, and they are
surely the basic reason to cast serious doubts on the formal $\beta$
detection from the marginalized likelihoods. The non--Gaussian
behavior is minimal for $\nu$ masses.


\begin{figure}[t!]
\begin{center}
\includegraphics[height=5.5cm,angle=0]{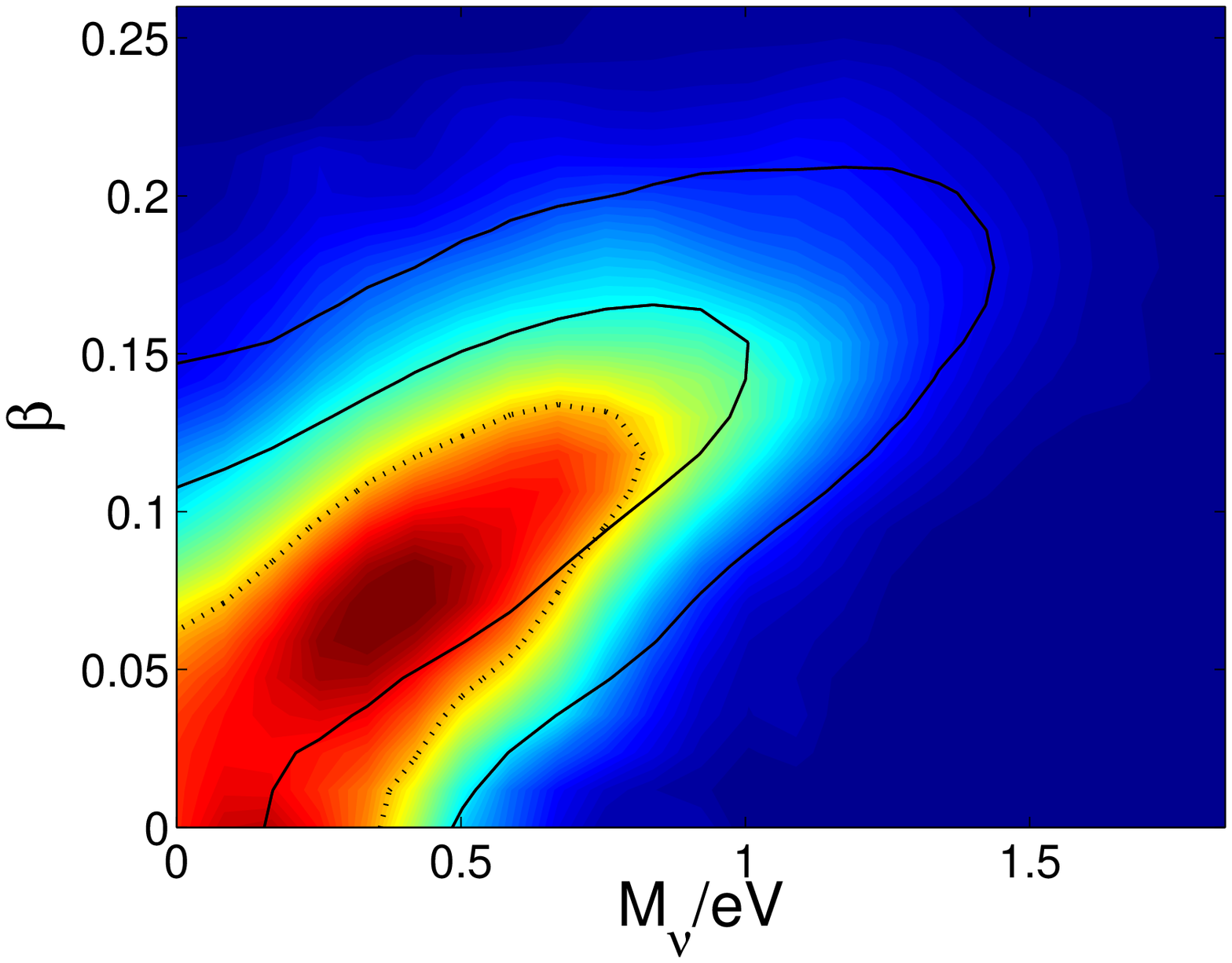}
\hskip.5truecm
\includegraphics[height=5.5cm,angle=0]{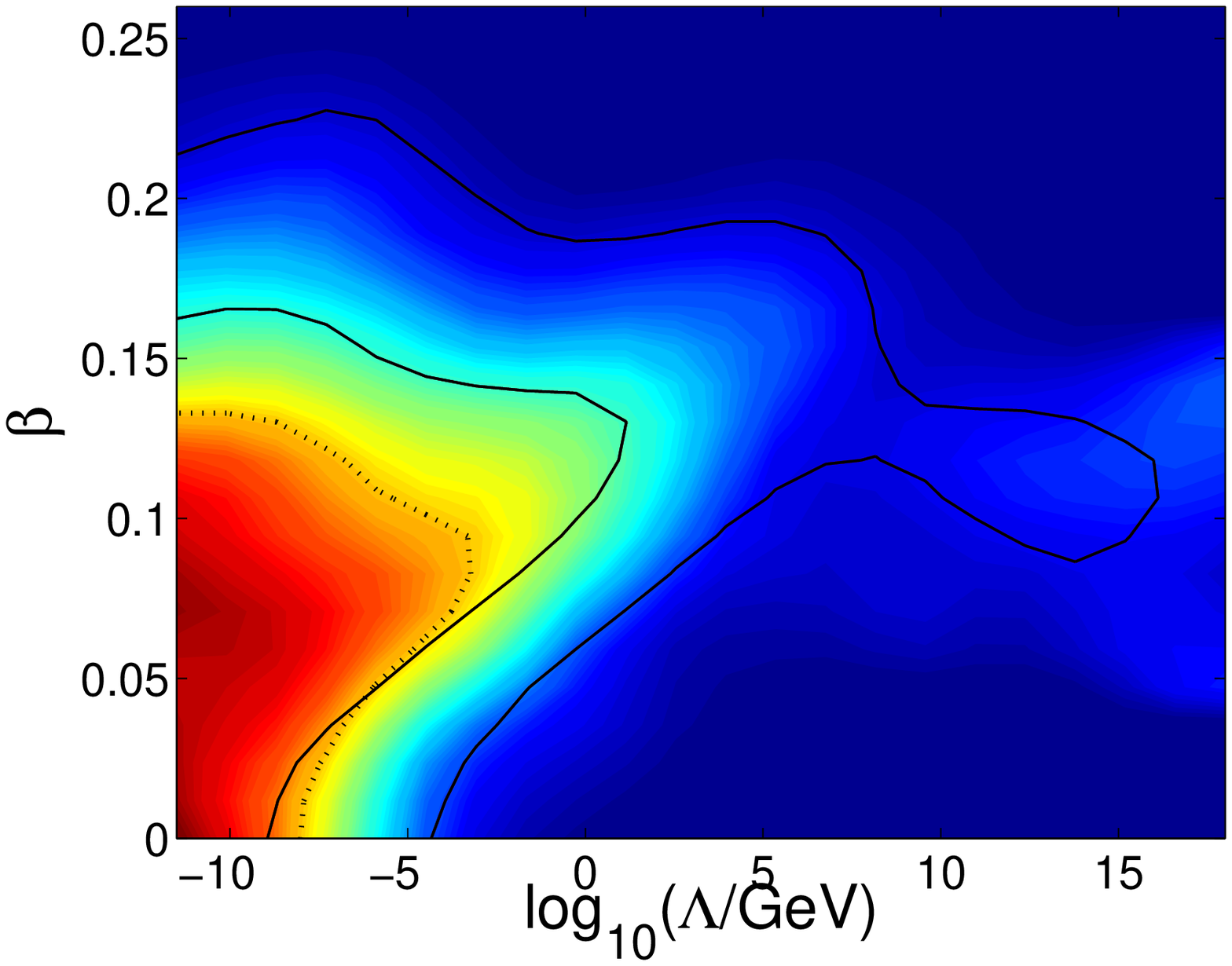}
\end{center}
\caption{Two parameter contours for the SUGRA model. Solid lines are
1-- and 2--$\sigma$ limits for marginalized likelihood.  Colors refer
to average likelihood, and the 50\% likelihood contour from the
average likelihood is indicated by the dotted line.}
\label{2DS}
\end{figure}

\begin{figure}[t!]
\begin{center}
\includegraphics[height=5.5cm,angle=0]{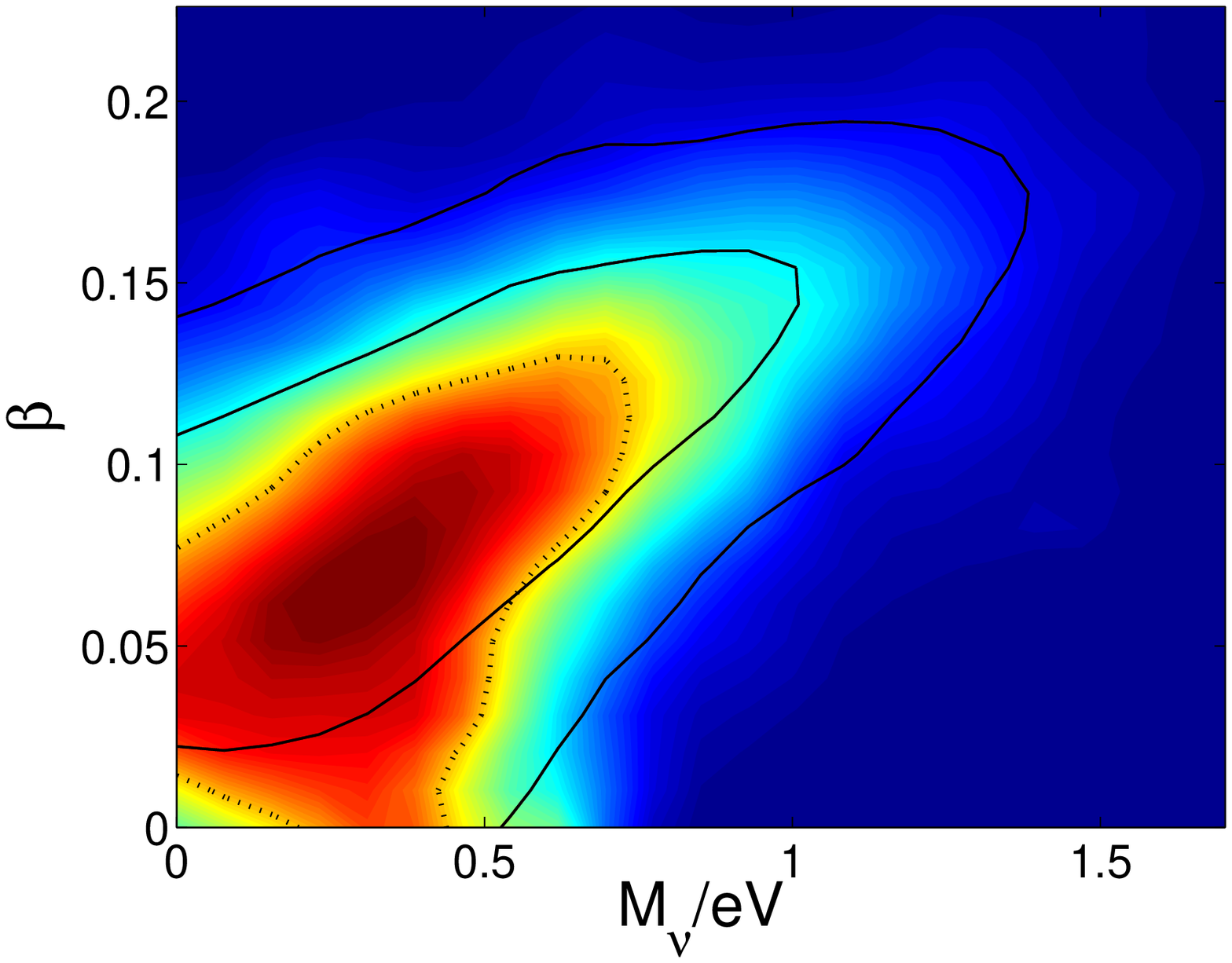}
\hskip.5truecm
\includegraphics[height=5.5cm,angle=0]{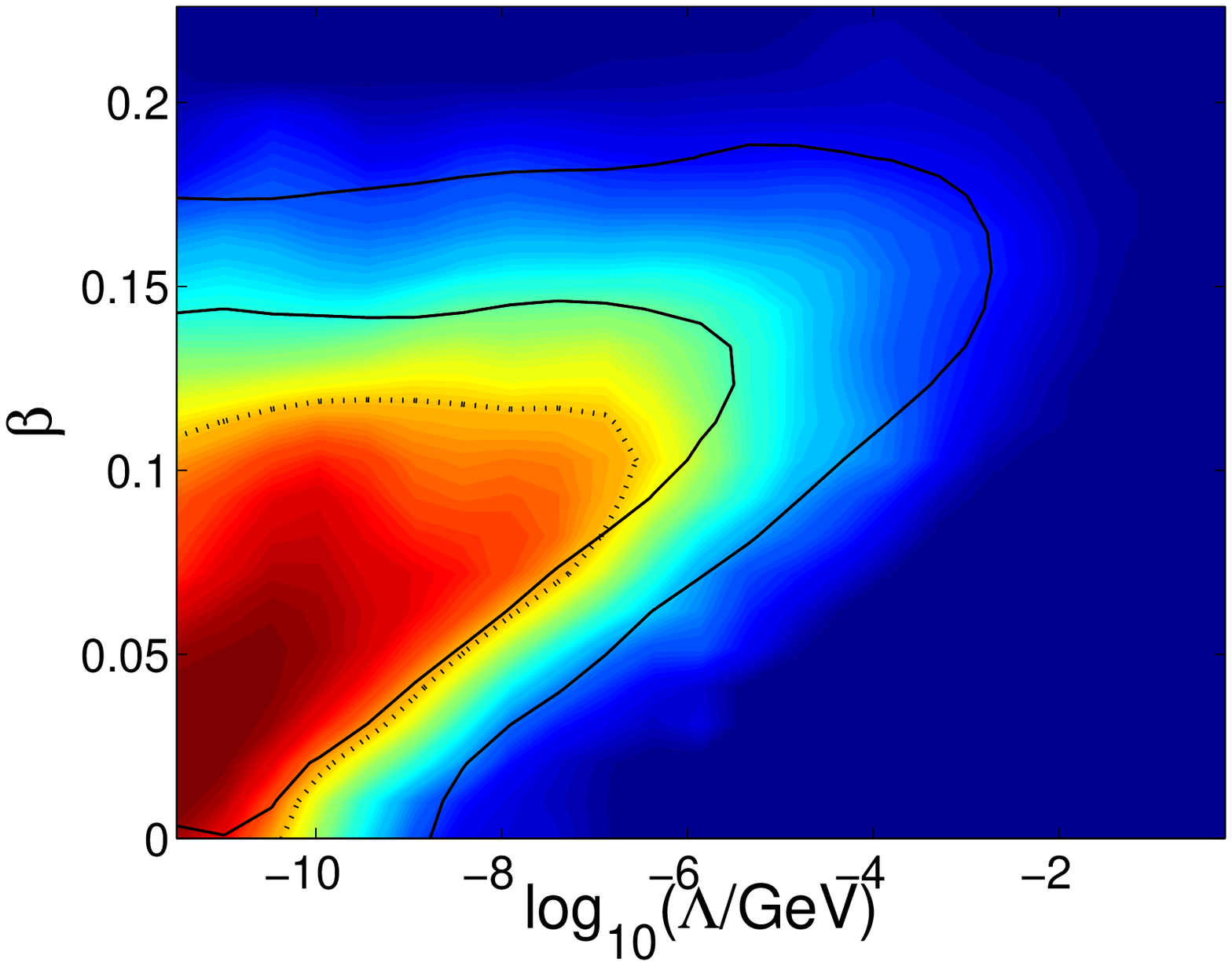}
\end{center}
\caption{As previous Figure, for a RP potential.}
\label{2DR}
\end{figure}

The most significant plots of this paper are Figures \ref{2DS} and
\ref{2DR}, concerning the SUGRA and RP potentials, respectively. In
these Figures, curves yield 1-- and 2--$\sigma$ contours for
marginalized likelihood distributions. On the contrary, colors yield
average likelihood distributions. The plots concern the $\beta$--\Mnu
~and the $\beta$--$\log_{10}(\Lambda/{\rm GeV})$ planes.

These plots allow, first of all, a better undestanding of the
discrepancy between mean and marginalized likelihood distributions.
Let us consider the correlations between $\beta$ and $\log_{10}
(\Lambda/{\rm GeV})$, shown in the lower panels. They indicate that
the same $\beta$ values are consistent with a fair range of
$\Lambda$'s. When marginalizing over the other parameters, $\Lambda$
included, this increases the weigth of high $\beta$ values.

In particular, Figure \ref{2DS} exhibits a peculiar high--$\Lambda$
tail: these 2--D plots show that, for suitable couplings $\beta \sim
0.1~$, models with $\Lambda$ up to $10^{15}\, $GeV (close to GUT) are
allowed within 2--$\sigma$'s. No such features is present for the RP
potential. Again, however, one appreciates that $\Lambda \sim $~MeV is
now allowed for RP if $\beta \sim 0.15$--0.17~.

Thus, when it comes to the fine--tuning paradox, the SUGRA potential
keeps a more satisfying solution than the RP potential, as a $\Lambda$
scale close to the EW scale or to the scale of the soft SUSY break is
consistent with data. Also RP, however, does no longer yield just
unacceptably low energy scales.

One should however also consider these potentials independently of the
supposedly underlying physics, as examples of rapidly or slowly
varying $w(a)$. Using such potentials we could actually inspect the
behaviors within these extremes, making recourse to a single parameter
$\Lambda$, instead of using, {\it e.g.}, the parametrization
$w_o$--$w'$, which requires 2 independent variables.

Let us then point out that, in the RP case, closer to $w = {\rm
const.}$, both the marginalized and the average likelihood
distributions on $\beta$ exhibit a maximum.

Also in the left panel of Figure \ref{2DR} the likelihood peak is
quite far away from $\beta=0$, $M_\nu = 0$. In the marginalized
likelihood distribution, such model is within the 95$\, \% $ C.L.. 

More in general, from the left panels of Figures \ref{2DS} and
\ref{2DR}, showing the 2D likelihood distributions in the $\beta$-\Mnu
~plane, we want to stress the following points:

\noindent
(i) The solid curves in the SUGRA case substantially overlap FM
results from \cite{lavacca} in the SUGRA case.

\noindent
(ii) Marginalized and average likelihood are different, as is expected
in the presence of a non--Gaussian behavior, but not significantly so.

\noindent
(iii) Top likelihood models are found for non--vanishing $\beta$ and
$M_\nu$; the best fit values are $\beta \sim 0.07$ and $M_\nu \sim
0.35\, $eV for both potentials.

\noindent
(iv) Models with $\beta$ up to 0.2 and $M_\nu $ up to 1.4 are allowed
with both potentials within the 95$\, \%$ C.L.$\, $. Again a result
apparently potential independent.

\noindent
(v) The limits on $\beta$ and \Mnu ~are strongly correlated, as
expected from the FM analysis \cite{lavacca}.

\noindent
(vi) According to the marginalized likelihood distributions,
$\beta=M_\nu=0$ models are allowed for both potentials. In the RP
case, however, they lay outside the 1--$\sigma$ limits.

\noindent
(vii) The fact that upper limits on $\nu $ masses are loosened by
almost a factor 2 arises from the degeneracy between $\beta$ and
\Mnu.

This confirms the results from \cite{lavacca} and shows that one
should be cautious when extracting neutrino mass limits from
cosmological observations, as they can heavily depend on the assumed
model range, as is the case here.

\section{Summary and conclusions}
\label{sec:summary}

One of the major discoveries of physical cosmology is the existence of
the dark sector of the Universe. Particles belonging to the {\it
standard model} of elementary interactions account for not more than
$5\, \%$ of its energy budget. Observations then go farther and show
that the remaining 95$\, \%$, which interacts with standard model
particles just through gravity, is not a single component, but needs
to be modeled at least through two independent fluids, with
drastically different state equations.

All existing data can be accomodated in a scheme assuming no energy
exchange between these two dark components. This is simple, but leads
to the well known {\it fine tuning} and {\it coincidence} paradoxes.

In this work we have explored  the option that, in the dark sector, energy
exchanges occur and may be described through a linear and constant
coupling. Clearly, this is just the next approximation beyond assuming
zero exchange and is however a phenomenological approach. Hopefully,
it may lead to constraints helping to single out a precise theory,
just as is done when the expression (\ref{wa}) is taken for the state
parameter scale dependence.

Such option was already explored in the past. The limits found for the
CDM--DE coupling were then rather deceiving. For $\beta$'s large
enough, the coincidence paradox could be significantly attenuated; but
the allowed $\beta$ range did not allow to go so far in this
direction.

The critical ingredient of this work amounts to considering
simultaneously coupling and $\nu$ masses. Massive $\nu$'s would be a
further component of the dark side, but unlike from CDM and DE,
however, they are particles already known from the standard model.
Furthermore, their role appears unessential to the formation of cosmic
structures, although their mass can actually modify the matter
distribution over the largest scales.
\begin{figure}[t]
\begin{center}
\includegraphics[height=8.6cm,width=10.truecm,angle=0]{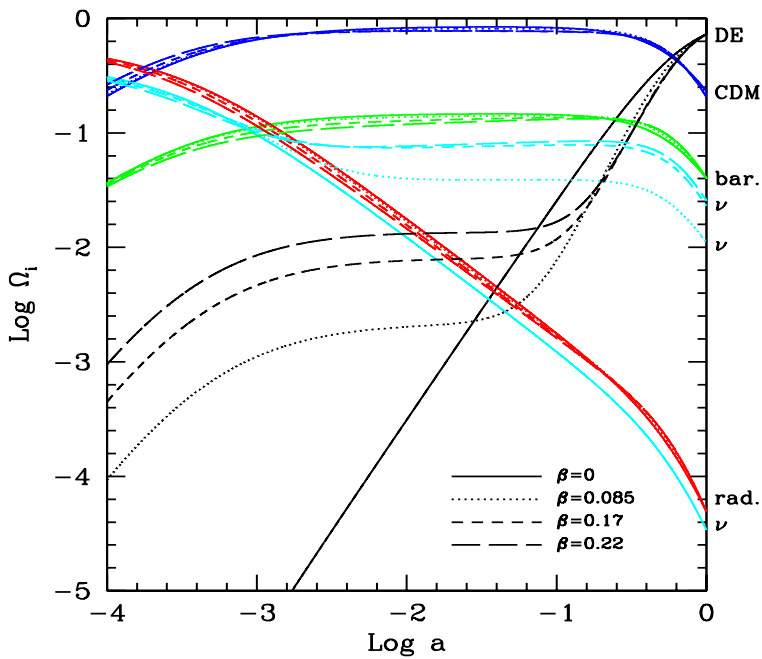}
\end{center}
\caption{Evolution of the density parameters in a SUGRA model with
coupling and $\nu$ mass. Colors refer to different components, as
specified in the frame; lines to the different models: \Mnu=0,
$\beta=0$ (continuous line); \Mnu=0.5~eV, $\beta = 0.085$ (dotted
line); \Mnu=1.1~eV, $\beta = 0.17$ (short dashed line); \Mnu=1.2~eV,
$\beta = 0.22$ (long dashed line). }
\label{RPom}
\end{figure}

When one allows simultaneously $\nu$ mass and $\beta$, we find models
with significantly higher likelihood. The upper limits on both
parameters are softened and increase by an essential factor $\sim
2$. This is a critical factor in both fields.  The new allowed \Mnu
~values approach the $\nu$--mass detection area in forthcoming tritium
decay experiments like KATRIN \cite{katrin}. Moreover, when
considering an external prior on $M_\nu$ from earth-based experiments,
the strong degeneracy between the coupling parameter $\beta$ and the
neutrino mass $M_\nu$ can be broken, gaining new insight into the DE
nature \cite{kristiansen09}.

Simultaneously, the new allowed $\beta$ values open the possibility of
a critically modified DE behavior. In Figure \ref{RPom} we show the
scale dependence of the cosmic components for various $\beta$--$M_\nu$
pairs. They tell us that models are allowed, within the 95$\, \%$
C.L., for which DE is still $\sim 1\, \%$ of the cosmic energy budget
up to $z \sim 10^3$; at higher redshifts it decreases just because the
photon-neutrino fraction increases, while we approach
matter--radiation equality. Such fraction attains 2--3$\, \%$ for the
most coupled case we considered in Fig.~\ref{RPom}, corresponding to
$C = 0.9/m_p$ with \Mnu=1.2~eV; this is at 2.3$\, \sigma$'s from the
best fit. Let us remind that $\Lambda$CDM cosmologies prescribe that,
at $z \sim 10^3$, DE bears less than $1: 10^{9}$ of the total energy
density.

Available data do not yet force us to require a non--vanishing CDM--DE
coupling. The statistical analysis of data still leads to an intricate
situation, where marginalized and average likelihoods exhibit
discrepancies. Furthermore, the likelihood distributions on the
coupling $\beta$ exhibits some dependence on the selected
self--interaction potential. Using a RP potential, the 0--0 option
appears rather unlikely, both through marginalized and average
likelihood distributions. A SUGRA potential, instead, yields a higher
likelihood for the 0--0 option. Accordingly, we believe that current
data do not allow a claim of ``$\beta$--detection'', while they
certainly allow to put upper limits to the coupling, which can be so
large, within the 95$\, \%$ C.L., to yield $C=1/2m_p$.

Furthermore, it is fascinating to notice that, if shortly forthcoming
particle data will set a lower limit to \Mnu, in the range they are
allowed to explore, this will imply an almost model independent
CDM--DE coupling detection, opening the way to a deeper understanding
of the dark sector of the Universe.

\acknowledgments 
JRK and RM acknowledge financial support from the Research Council of
Norway. LPLC was supported by NASA grant NNX07AH59G for this work.

\end{document}